\documentclass[preprint,12pt]{aastex}
\def\lap{\lower.5ex\hbox{$\; \buildrel < \over \sim \;$}}
\def\gap{\lower.5ex\hbox{$\; \buildrel > \over \sim \;$}}

\def \eg{{e.g.,}}

\def \etal{{et~al.\null}}
\def \ie{{i.e.,}}

\title{Virgo's Intracluster Globular Clusters as Seen by the {\sl
Advanced Camera for Surveys}}

\begin{document}

\shorttitle{Intracluster Globular Clusters}

\author{Benjamin F. Williams, Robin Ciardullo}

\affil{Department of Astronomy \& Astrophysics, The Pennsylvania State
University \\ 525 Davey Lab, University Park, PA 16802}

\email{rbc@astro.psu.edu, bwilliams@astro.psu.edu}

\author{Patrick R. Durrell, John J. Feldmeier}

\affil{Department of Physics \& Astronomy, Youngstown State University,
Youngstown, OH 44555}

\email{prdurrell@ysu.edu, jjfeldmeier@ysu.edu}

\author{Steinn Sigurdsson, Matt Vinciguerra}

\affil{Department of Astronomy \& Astrophysics, The Pennsylvania State
University \\ 525 Davey Lab, University Park, PA 16802}

\email{steinn@astro.psu.edu, mattv@astro.psu.edu}

\author{George H. Jacoby}
\affil{WIYN Observatory, P.O. Box 26732, Tucson, AZ 85726}

\email{jacoby@wiyn.org}

\author{Ted von Hippel}

\affil{The University of Texas, Department of Astronomy, 
1 University Station C1400, Austin, Texas 78712}

\email{ted@astro.as.utexas.edu}

\author{Henry C. Ferguson}

\affil{Space Telescope Science Institute, 3700 San Martin Drive, 
Baltimore, MD 21218}

\email{ferguson@stsci.edu}

\author{Nial R. Tanvir}

\affil{Department of Physics and Astronomy, University of Leicester,
Leicester, LE1 7RH, UK}

\email{nrt@star.herts.ac.uk}

\author{Magda Arnaboldi}

\affil{European Southern Observatory, Karl-Schwarzchild-Str.~2, 
85748 Garching, Germany}

\email{marnabol@eso.org}

\author{Ortwin Gerhard}

\affil{Max-Planck-Institut fuer Extraterrestrische Physik, 
P.O. Box 1312, D-85741 Garching, Germany} 

\email{gerhard@mpe.mpg.de}

\author{J. Alfonso L. Aguerri}

\affil{Instituto de Astrof\'isica de Canarias, 
via L\'actea, s/n, 38205, La Laguna, Tenerife, Spain}

\email{jalfonso@ll.iac.es} 

\and

\author{Ken C. Freeman}

\affil{Mount Stromlo Observatory, Research School of Astronomy and 
Astrophysics, Mount Stromlo Observatory \\
The Australian National University, ACT 0200 Australia}

\email{kcf@mso.anu.edu.au}

\begin{abstract}

We report the discovery of 4 candidate intracluster globular clusters
(IGCs) in a single deep {\sl HST ACS\/} field of the Virgo Cluster.
We show that each cluster is roughly spherical, has a magnitude near
the peak of the Virgo globular cluster luminosity function, has a
radial profile that is best-fit by a King model, and is surrounded by
an excess of point sources which have the colors and magnitudes of
cluster red giant stars.  Despite the fact that two of our IGC
candidates have integrated colors redder than the mean of the M87
globular cluster system, we propose that all of the objects are
metal-poor with [M/H] $< -1$.  We show that the tidal radii of our
intracluster globulars are all larger than the mean for Milky Way
clusters, and suggest that the clusters have undergone less tidal
stress than their Galactic counterparts.  Finally, we normalize our
globular cluster observations to the luminosity of intracluster stars,
and derive a value of $S_N \sim 6$ for the specific frequency of Virgo
intracluster globular clusters.  We use these data to constrain the
origins of Virgo's intracluster population, and suggest that globular
clusters in our intracluster field have a different origin than
globular clusters in the vicinity of M87.  In particular, we argue
that dwarf elliptical galaxies may be an important source of
intracluster stars.

\end{abstract}


\section{Introduction}

The constituents of intracluster space can tell us a great deal about
the history of galaxies and clusters.  As a cluster forms, tidal
interactions between galaxies and with the cluster potential affect
the internal structure of galaxies, altering both their morphological
and photometric properties \citep[][and many others]{butcher1978,
dressler1980, goto2003, coenda2006}.  At the same, these interactions
also liberate material into intergalactic space, thus creating a
fossil record of the encounters.  By studying the composition,
distribution, and kinematics of these orphaned objects, we can examine
the physics of tidal stripping, the distribution of matter in and
around galaxies, and the initial conditions and history of cluster
formation \citep[][and many others]{merritt1984, west95, gregg1998,
sommer2005}.

Intracluster globular clusters (IGCs) are an especially useful probe
of these processes \citep{west95}.  As a globular cluster evolves, it
preserves information about the time of its creation, the chemistry of
the gas out of which it formed, and even the gravitational forces to
which it has been exposed \citep[see][and references
therein]{ashman98}.  Consequently, a large sample of IGCs can be used
to trace the history of galaxy interactions and constrain both the
epoch of cluster formation and the system's dynamical history.

Unfortunately, collecting and measuring a large sample of intracluster
globular clusters is difficult.  Globular clusters in the halos of
galaxies are routinely identified as an excess of point sources above
the background, and searches for such objects have been conducted in
$\sim$100 systems out to $\sim$100~Mpc \citep[\eg][]{harris79,
kisslerpatig, kundu2002}.  However, as galactocentric distances
increase, the surface densities of clusters decrease, so in
intracluster space, the identification of globular clusters as point
sources is exceedingly difficult.  As a result, there have been only a
few, mostly indirect, studies of IGCs \citep{west03, jordan03,
franch03, bassino03}, and their use as cosmological probes has largely
been unexploited.

Here we describe the results of a {\sl Hubble Space Telescope\/ (HST)}
search for intracluster globular clusters (IGCs) in the nearby Virgo
Cluster.  At our adopted Virgo distance \citep[16.2~Mpc, see
discussion in][]{VICS1}, globular clusters have half-light radii of
$\gap~0 \farcs 05$, allowing them to be resolved on images taken with
the Advanced Camera for Surveys ({\sl ACS\/}).  Moreover, because of
the {\sl ACS'} excellent sensitivity, it is possible to use the
instrument to detect individual stars within the clusters and estimate
their metallicities via the color of the red giant branch.  In
Section~2, we describe our survey, and announce the discovery of four,
well-resolved IGC candidates in Virgo.  In Section~3, we discuss the
metallicities of these objects, and show that all are metal poor, with
photometric properties that differentiate them from the globular
clusters of Virgo's central cD galaxy, M87.  In Section~4, we compare
the candidate IGCs to Galactic globular clusters and show that their
half-light and tidal radii are larger than their Milky Way
counterparts.  We attribute these properties to the IGCs' lack of
tidal processing, and use the radii to constrain the clusters'
origins.  We conclude by estimating the specific frequency of globular
clusters in Virgo's intracluster space, and discussing the
implications this number has for the origin of IGCs and future IGC
surveys.

\section{Observations and Reductions}

Between 30 May 2005 and 7 June 2005 we used the {\sl Advanced Camera
for Surveys\/} on the {\sl Hubble Space Telescope\/} to obtain deep
F606W and F814W images of a single Virgo intracluster field
($\alpha(2000) = $12:28:10.80, $\delta(2000) = $12:33:20.0,
orientation 112.58 degrees), $\sim 0.67$~deg ($\sim 200$~kpc) from any
large galaxy.  The F814W ($I$-band) data consisted of 22 exposures
totaling 26880~s of integration time; the F606W (wide $V$-band)
observations included 52 exposures totaling 63440~s.  These data were
co-added using the {\tt multidrizzle} task within PyRAF
\citep{koekemoer}: this procedure removed all the cosmic rays,
corrected the instrument's geometric distortions, and improved the
sampling of our data to $0\farcs 03$~pixel$^{-1}$.  The details of
these reductions, and an image of the field illustrating its position
in the cluster is given by \citet{VICS1}.

After combining the data, we used SEextractor \citep{bertin96} to
identify all sources (extended and unresolved) brighter than $m_{\rm
F814W} = 24.5$, near the peak of the globular cluster luminosity
function in Virgo \citep[$V=23.7$,][]{whitmore95}.  We then
cross-correlated this list with the point source identifications from
DAOPHOT~II \citep{stetson90}, and searched for objects surrounded by a
statistical excess of stars ($N \ge 4$).  This procedure yielded eight
sources, three of which were obvious background galaxies with internal
structure, and one of which was clearly surrounded by misidentified
background galaxies.  However the remaining four objects had the
properties expected for IGCs.  Each was well-resolved and roughly
spherical ($b/a \geq 0.88$), each had an integrated F814W magnitude
near the peak of the Virgo globular cluster luminosity function, and
each was surrounded by point sources which had the colors and
magnitudes of Virgo Cluster red giant stars.  The coordinates of these
four sources, aligned to the astrometry of the automatic plate
measuring (APM) machine catalog,\footnote{see
http://www.ast.cam.ac.uk/$\sim$apmcat/} are given in Table~\ref{tab1};
images of the objects are displayed in Figure~\ref{pics}.

We performed photometry on the point sources surrounding the IGC
candidates using DAOPHOT~II \citep{stetson90}.  Instrumental
magnitudes in $0\farcs 5$ apertures were computed via
point-spread-function fitting.  These magnitudes were then
extrapolated to infinite apertures and placed on the Vega magnitude
system using the correction parameters and photometric zero points
given by \citet{sirianni05}.  Photometry of the candidate IGCs
themselves was performed with the IRAF\footnote{IRAF is distributed by
the National Optical Astronomy Observatory, which is operated by the
Association of Universities for Research in Astronomy, Inc., under
cooperative agreement with the National Science Foundation.} task {\tt
phot} using a series of concentric circular apertures, ranging in size
from $0\farcs 0225$ to $1 \farcs 35$.  Within $0 \farcs 25$ these
aperture radii were incremented in $0\farcs 0225$ intervals; outside
this radius, aperture widths were increased to maintain a near
constant photometric error.  Again, these instrumental magnitudes were
converted to the Vega magnitude system using the zero points of
\citet{sirianni05}.  Transformations to the standard $VI$ magnitude
system were performed using the coefficients of \citet{rejkuba}.  We
note, however, that these transformations have a rather large scatter,
$\sim 0.05$~mag.  Consequently, this last step was only used to
compare the integrated magnitudes of our globular cluster candidates
with other measurements in the literature.  Whenever possible, we
confined our analysis to the F606W$-$F814W magnitude system.

To translate our photometric measurements into physical parameters, we
first dereddened the observed magnitudes using the
\citet{schlegel1998} value for Galactic foreground extinction ($E(B-V)
= 0.025$).  With the reddening law of \citet{cardelli1989} and the
{\sl ACS\/} filter transformations of \citet{sirianni05}, this
corresponds to $A_V=0.077, A_I=0.046, A_{F606W}=0.069,$ and
$A_{F814W}=0.045$.  Total luminosities for the IGC candidates were
then calculated assuming a Virgo distance of 16.2~Mpc, and a
bolometric correction of $BC_V \approx -0.5$ (with $M_{bol_{\odot}} =
4.74$).  Finally, these luminosities were used to estimate masses by
assuming a mass-to-light ratio of $M/L_V = 2.3$, which is typical for
Galactic clusters \citep{pryor93}.

To test whether our candidate IGCs are true globular clusters, we
convolved a series of \citet{king62} model profiles with a
\citet{moffat} representation of the F814W filter
point-spread-function, and fitted the resultant curve to the objects'
radial profiles using a $\chi^2$ minimization procedure.  The best
fits are shown in Figure~\ref{king}, with the photometric errors
increased by 5\% (added in quadrature) to account both for deviations
between the true PSF and our Moffat function, and for the ``red halo
effect'' \citep{sirianni05}.  The best-fit core radii, tidal radii,
and half-light radii are given in Table~\ref{tab1}; the errors on
these numbers are the standard deviations of fits to a series of Monte
Carlo simulations of each brightness profile.  (The data for IGC-3
(our faintest candidate) was not of sufficient precision to constrain
the tidal radius, so no errors are given for this object.)  In all
four cases, our convolved King profiles provide good fits to the data,
with $\chi^2/\nu \leq 1.5$.  This contrasts with fits which use the
\citet{deV} $r^{1/4}$-law or an exponential disk (also given in
Table~\ref{tab1}), which generally give poorer $\chi^2$ values.  The
quality of the fits strongly supports the conclusion that these
objects are, indeed, globular clusters.

Additional support for the globular cluster interpretation comes from
the point sources surrounding each IGC candidate.  The mean density of
all unresolved sources detected in our intracluster survey field (down
to a limiting magnitude of $m_{\rm F814W} = 28.5$) is
480~arcmin$^{-2}$ \citep{VICS1}.  Thus, we would expect $\sim 0.67$
stars to be projected between $0 \farcs 3$ and $1 \farcs 3$ from the
center of each IGC candidate.  Even the faintest of our candidates has
6 times this number; the Poisson probability of having four or more
stars randomly projected about a $m_{\rm F814W} < 24.5$ object in our
field is less than 0.5\%.  Since there are 195 such objects present in
the region, at most one of our IGC candidates is expected to be a
chance superposition of stars around a bright unrelated object, and a
visual examination of the candidates reduces this number, further.

The final piece of evidence supporting of IGC identification is the
magnitude of the brightest stars surrounding each cluster, $I \sim
27$.  This is the magnitude expected of red giant stars at the
distance of Virgo \citep{ferguson1998,durrell02}.  It therefore
seems likely that the sources surrounding each IGC candidate are red
giants bound to the clusters.

Can the objects be background galaxies or some other source unrelated
to globular clusters?  The radial profile of IGC-1 can be fit with
both a King model and an $r^{1/4}$-law ($\chi^2/\nu = 1.5$), so it is
conceivable that this object is a field elliptical galaxy.  However,
IGC-1 is also bright enough to have photometry from the Sloan Digital
Sky Survey \citep{sloan4}, and its Sloan $u' - g'$ and $r' - i'$
colors ($0.3 \pm 0.7$ and $0.0\pm 0.4$, respectively) are bluer than
those of any normal elliptical at any redshift \citep{csabai2003}.
This fact, along with the object's $g' - i'$ and $V-I$ colors, which
are bluer than those of local small ellipticals \citep{csabai2003,
fukugita1995} but similar to those of Virgo globulars
\citep[\eg][]{forbes2004, kundu1999}, makes it extremely unlikely that
the object is a background galaxy.  Similarly, although the radial
profile of our faintest candidate, IGC-3, {\it can\/} be fit with an
$r^{1/4}$-law (with $\chi^2/\nu = 0.3$), King models or a
\citet{sersic} profile with $n = 0.5$ (\ie\ an isothermal
distribution) generate an even lower $\chi^2$, and the object's $V-I$
color is again much bluer than that expected from a normal elliptical.
Finally, IGC-2 and IGC-3 are our most elongated IGC candidates (see
Table~\ref{tab1}), with shapes that are more eccentric than $\sim
80\%$ of Galactic globulars \citep{gccatalog}.  While, it is possible
the more elliptical Galactic clusters are tidally disturbed by the
Galaxy, it is clear that these ellipticities do not rule out a
globular cluster classification.  In any case, King profiles provide
much better fits to these candidates than any exponential or $r^{1/4}$
law.

In fact, the only objects that could reproduce the observed properties
of our IGC candidates are the nuclear remains of tidally stripped
dwarf galaxies.  Such an origin has been proposed for the Milky Way
object $\omega$~Cen \citep[\eg][]{bekki2003,mackey2005}, the giant
globular cluster G1 in M31 \citep{meylan2001}, the most massive
clusters of NGC~5128 \citep{martini2004}, and the ultra-compact dwarfs
of Virgo and Fornax \citep{drinkwater00, drinkwater04, jones2006}.
However, all of our IGC candidates are much fainter than these unusual
objects: for example, $\omega$~Cen is over a magnitude brighter than
our most luminous IGC candidate, and the Ultra-compact dwarf galaxies
found by \citet{drinkwater00} and \citet{jones2006} are brighter
still.  Of course, it is difficult to completely exclude the
possibility that our IGCs are stripped dwarfs: $\omega$ Cen is more
than a thousand times closer than Virgo, and its classification is
still controversial \citep{vanleeuwen2002}.  Nevertheless, given that
all four candidates have luminosities near the peak of the globular
cluster luminosity function, the simpliest explanation for these
sources is that they are, indeed, normal globular clusters at the
distance of Virgo.

\section{The Metallicities of the IGCs}
Figure~\ref{cmds} plots the point sources within $1\farcs 3$~arcsec
($\sim$100 pc at Virgo) of each globular cluster candidate on the
F606W$-$F814W color-magnitude diagram.  Overplotted are isochrones for
an old (12.5~Gyr) stellar population \citep{girardi02, girardi06} at
the distance of Virgo.  These sparse CMDs suggest that all four of our
IGC candidates are metal-poor.  The most metal-rich of the group,
IGC-4, has [M/H] $\sim -1.3$, while the stars of IGC-2 and IGC-3 fall
close to most metal-poor isochrone ([M/H] $\sim -2.3$).  Curiously,
four of the point sources surrounding IGC-1 fall blueward of any of
the \citet{girardi06} isochrones, in a region of the HR diagram
occupied primarily by background sources \citep{VICS1}.  It is
possible that either there is an overdensity of background objects in
this part of the field or that the effects of crowding have produced
errors in the stellar photometry larger than the standard errors shown
on the CMD.  In any case, if we exclude these objects from the
analysis, the remaining stars of the cluster imply a metallicity of
[M/H] $\sim -1.5$.

The metal-poor nature of our globular cluster candidates is supported
by their integrated colors, though not to the extent that one might
expect.  In most large galaxies, the distribution of globular cluster
colors is bimodal \citep[\eg][]{gkp99,lar01,kw01,harris06}.  For
example, the color distribution of M87 globulars is well-modeled by
two Gaussians, one with a peak at $V-I = 0.95$ and the other centered
at $V-I = 1.20$ \citep{whitmore95, kundu1999}.  This division has
generally been interpreted as evidence for the existence of two
separate populations of clusters, one consisting of blue,
``metal-poor'' objects, and the other comprised of ``metal-rich''
systems \citep[][but see \citet{yoon06} for an alternative
explanation]{harris06}.  The integrated colors of clusters IGC-3 and
IGC-4 clearly place them in the metal-poor category, as one might
expect from the colors of their halo stars.  However, IGC-1 and IGC-2
both have colors that fall slightly to the red of the dividing line.
This seems incompatible with the colors of the systems' red giant
stars.

To investigate this inconsistency, we measured the objects' radial
color profiles.  As Figure~\ref{colorprof} demonstrates, IGC-1 and
IGC-2 both have significant color gradients, with the clusters'
interiors being redder than their halos by $\sim 0.2$~mag.  Such
gradients are usually associated with galaxies, and, as mentioned
above, it is conceivable that these two clusters are actually the
stripped remains of compact dwarf galaxies \citep{drinkwater04,
jones2006}.  But this need not be the case: in the Galaxy, one third
of all globular clusters have similar red-to-blue gradients
\citep{chun79, sohn98}.  Whether these gradients are caused by the
effects of mass segregation, the random presence of a few relatively
bright stars \citep{peterson86}, or chemical inhomogeneities in the
stellar populations \citep{freeman80} is unclear.  However, it does
explain how our two IGC candidates can have neutral colors, but still
exhibit metal-poor CMDs.  It is therefore possible that these two IGCs
are of intermediate metallicity.  Future spectroscopy can provide a
definitive answer to this question.

The fact that all four of our IGC candidates are relatively blue
stands in marked contrast to the color distribution of clusters in
M87's inner regions.  M87's globulars have a mean $V-I = 1.10$ and
$\sim 60\%$ are classified as ``metal-rich'' based on their red
colors.  All of our IGC candidates have metal-poor RGBs, and our two
reddest clusters just barely fall on the red-side of the color
distribution.  To a limited extent, this is consistent with the
results of \citet{harris06}, who showed that outside of $\sim 5$~kpc,
the ratio of red to blue clusters surrounding brightest cluster
galaxies drops dramatically.  However, in the \citet{harris06} sample,
the fraction of red clusters never drops below $\sim 40\%$, even at
galactocentric distances of $\sim 30$~kpc.  The IGCs in our
intracluster field are $\sim 200$~kpc from any galaxy, and the
probability of observing four clusters with $(V-I) \lesssim 1.16$ out
of the \citet{harris06} distribution is just $\sim 10\%$.  These
numbers suggests that the IGC population in our field is fundamentally
different from that associated with brightest cluster galaxies.

Alternatively, it is possible that the four globular cluster
candidates observed in our small intracluster field are not
representative of the Virgo IGC population as a whole.  Three of the
four clusters, IGC-1, 2, and 3, are located in a line which runs
north-south along the western half of our field.  Since these clusters
are also the most metal-poor of our candidates, it is possible that
all three originated in a single stripped galaxy, and that the stream
has not yet completely mixed with the general intracluster population.
Indeed, the spatial sub-structure exhibited by the metal-poor stars in
our {\sl ACS\/} field is evidence for just such a scenario
\citep{VICS1}.  If coherent streams are common, then a much wider
survey will be needed to reliably measure the properties of Virgo's
globular clusters.

\section{The IGC Radial Profiles}

Figure~\ref{rcrt} compares the core and tidal radii of our IGC-1, 2, and 4
with those of Galactic globular clusters, using our adopted Virgo distance.
(IGC-3 is not plotted, since its tidal radius is unconstrainted.)
From the figure, we can see that, though the core radii of
the two sets of objects are comparable, the tidal radii of the
intergalactic objects are larger than most of their Milky Way
counterparts.   This is not a selection effect:  since our IGC search
criteria included all sources detected by SExtractor, our sample is not
biased by size.   In fact, a close examination of Figure~\ref{rcrt}
shows that the tidal radii of our IGC candidates are similar to those
of the Milky Way clusters with large Galactocentric distances.  This
agrees with the thesis that globular clusters inside
large galaxies are continually affected by tidal stress. 

Such stress is thought to play a key role in the evolution of
galaxy-bound clusters.  For example, analyses by \citet{aguilar1988}
and \citet{fall01} suggest that over a Hubble time, a large fraction
of globular clusters within a Milky Way-type spiral will be either
stripped or destroyed.  However, if our IGCs were created {\it in
situ,} or if their parent galaxies were low-mass objects, then tides
have never been important for their dynamical evolution.  Support for
this idea also comes from the observations of \citet{jordan03}, who
showed that in the rich Abell Cluster~1185, the half-light radii of
globular clusters systematically increases with galactocentric
distance.

The large tidal radii of the clusters provide a hint about the length
of time for which the IGCs have been free-floating.  Dynamical models
predict that globular clusters recover from galactic tidal shocking on
a half-mass relaxation timescale, which is typically 5--10 $\times
10^8$ yr \citep{johnston1999}.  After several of these relaxation
times, the clusters will lose any structures caused by past tidal
effects and approach a distribution governed by the gradient of the
galaxy cluster's potential, a value that is $\sim 10^{-3}$ times less.
The fact that the IGCs are well-fitted by King profiles with finite
tidal radii suggests our objects were once affected by the tidal field
of a galaxy \citep{heggie2001}, but are now internally evolving
towards a state with little tidal truncation.  Since this process can
take 5-10 Gyr \citep{johnston1999}, the observations imply that these
IGCs have been free-floating, unaffected by strong tidal influences,
for several Gyr.

\section{Origins of the Clusters} 

In order to place these IGC candidates into a context of galaxy
cluster evolution, it is important to compare our surface density
results with previous surveys for these objects.  The existence of
four IGCs within our 11.4~arcmin$^{-2}$ field implies that Virgo's IGC
surface density is $\sim 10^{-4}$~arcsec$^{-2}$.  If we scale this
number to the distance of Coma \citep[a distance ratio of
6;][]{dressler1984}, then our data imply a surface density that is
safely below the upper-limit of 0.004~arcsec$^{-2}$ measured by
\citet{franch03}.  In contrast, a scaling of our surface density to
the distance of A1185 ($cz = 9800$ km s$^{-1}$) yields a value that is
a factor of two larger than that observed by \citet{jordan03}.
However, since their survey only reached one magnitude {\it brighter}
than the peak of the globular cluster luminosity function, while our
observations go $0.8$~mag {\it fainter} than the peak, the two values
are compatible.

The number of IGCs places an interesting constraint on the specific
frequency ($S_N$) of globular clusters in Virgo's intracluster
environment.  Star counts (after the statistical removal of unresolved
background galaxies) in our $11.4$~arcmin$^2$ field reveal $\sim 5300$
stars brighter than $m_{\rm F814W} \sim 29$ \citep{VICS1}.  If we
extrapolate these counts down to the main sequence using the
isochrones of \citet{girardi06}, then the data imply an intracluster
surface brightness brightness of $\mu_V \sim 28.1$~mag~arcsec$^{-2}$
\citep[for details, see][]{VICS1}, an absolute total luminosity of
$\sim$2$\times 10^5~{\rm L}_{\odot}$~kpc$^{-2}$, and a globular
cluster specific frequency $S_N \sim 6$.  This relatively high value
suggests that we have not missed a significant number of IGCs, and
that the properties of these four IGCs may be representative of the
IGC population in Virgo.

Our value of $S_N \sim 6$ can be used to place a constraint on the
origins of the Virgo intracluster population.  If most of the stars in
Virgo's intergalactic space originated in typical spirals \citep[\ie\
were liberated via the galaxy harassment scenario of][]{moore96,
moore99}, then we would expect to measure a much lower value for the
globular cluster specific frequency, with $S_N \sim 1$
\citep[\eg][]{goudfrooij03, chandar04}.  This argument may apply to
low surface brightness spirals as well (J.H. Kim et al., in
preparation).  Alternatively, if intracluster stars are the disrupted
remains of low-luminosity dwarfs, then $S_N$ should be $\sim 20$
\citep{grebel00, strader03}.  Our value for the specific frequency of
globular clusters lies between these two extremes, in the range
normally associated with cD galaxies \citep{west95, forbes97} and
dwarf ellipticals \citep{durrell96, miller98, beasley06}.  In fact,
dwarf elliptical galaxies are the most common morphological type in
galaxy clusters \citep{binggeli88}, and the Virgo core currently
contains about 900 of these objects.  This is a significant number of
galaxies: \citet{durrell02} estimate that if two-thirds of all dwarf
ellipticals are destroyed through gravitational interactions over a
Hubble time, then their remnants could account for Virgo's entire
intracluster population.

The preceding has two caveats.  First, the low metallicities and high
specific frequency of IGCs could be due to preferential stripping of
globular clusters during tidal encounters.  The radial distribution of
clusters within a galaxy is often flatter than that of the galaxy's
light \citep[\eg][]{puzia2004, forbes2006}; this fact is consistent
with the idea that such systems are often formed during galactic
mergers and interactions \citep{ashman1992}.  Since clusters (and
stars) in the outskirts of galaxies are more susceptible to tidal
forces than interior objects, this mismatch can lead to the increased
stripping of globular clusters with respect to the stars.  The result
is that in rich clusters, where tidal encounters are important, the
specific frequency of clusters in intergalactic space can be enhanced
at the expense of galactic values.  There is some evidence for just
this effect: in Virgo, some galaxies have lower values of $S_N$ than
their field counterparts \citep[see][and references
therein]{elmegreen1999}.  Moreover, since systems of blue globular
clusters often have flatter radial distributions than those of red
clusters \citep[\eg][]{bassino2006, forbes2006, harris06}, this
process can explain why the four IGC candidates found in our survey
are all metal-poor.

A second caveat to our measurement of $S_N$ concerns the survival of
clusters in the galactic and extragalactic environments.  In large
galaxies, bulge shocks, disk shocks, and dynamical friction all take
their toll on the globular cluster population, so that, over a Hubble
time, a large fraction of clusters will be destroyed
\citep{aguilar1988}.  Such processes do not occur in intergalactic
space.  Consequently, while values of $S_N$ in a passively evolving
galaxy can decline with time, the specific frequency of IGCs can
actually increase (This is because the IGCs are not destroyed and the
luminosity of the normalizing intracluster stellar population
decreases as it ages).  The importance of this effect is difficult to
model, since it depends critically on where and when the IGCs
originally formed.  However, like the effects of preferentially
stripping, this mechanism will produce higher values of $S_N$ for the
intracluster environment than for the clusters' parent galaxies.

The metallicities of our IGC candidates alone do not help determine
the clusters' origins, as most globular cluster systems contain a
significant blue (metal-poor) component.  However, the relatively blue
colors and metal-poor CMDs of our candidates do support the hypothesis
of \citet{harris06} that redder, more metal-rich clusters must form
later in the deeper potential wells of major galaxies.  The fact that
none of our IGCs are metal-rich implies that, once formed, it is
difficult to eject red clusters into intergalactic space.

\section{Future Possibilities}
In order to properly investigate the systematics of IGCs, one needs a
much larger sample of objects.  The key to obtaining such a sample is
spatial resolution: with ground-based images, it is extremely
difficult to distinguish IGCs from the (much more numerous) background
contaminants.  However, the fact that four IGCs were discovered in a
single {\sl ACS\/} field suggests that with a few additional {\sl
HST\/} pointings, one can identify an astrophysically interesting
sample of such objects.  With {\sl HST\/} resolution, one can obtain
the IGCs' structural parameters, measure their tidal radii, and
investigate the systematics of a set of clusters that have evolved in
a largely stress-free environment.  Moreover, with follow-up
ground-based spectroscopy one can measure conclusive ages and
metallicities.

Such a sample can be a powerful probe of galactic and cluster
evolution.  By comparing the luminosities of globular clusters in and
outside of galaxies, one can test for the effects that bulge and disk
shocking have on the globular cluster luminosity function.  Similarly,
by examining the distribution of tidal radii for IGCs, one can probe
the length of time these objects have been in the intergalactic
environment, and complement population constraints imposed by the
intracluster stars.  Finally, with a large sample of clusters, we can
test whether the bimodal color and metallicity distributions often
seen around large galaxies extend to the intracluster population, and
whether ``red'' clusters can be liberated from their parent galaxies.
Such tests, in turn, can place new constraints on the formation of
these objects.

Support for this work was provided by NASA through grant number
GO-10131 from the Space Telescope Science Institute and by NASA through
grant NAG5-9377.


\clearpage

\begin{deluxetable}{lcccc}
\tablewidth{0pt}
\tablecaption{Measured parameters for IGC candidates}
\tablehead{
\colhead{Property} 
&\colhead{IGC-1}
&\colhead{IGC-2}
&\colhead{IGC-3}
&\colhead{IGC-4}
}
\startdata
$\alpha(2000)$ &12:28:04.78 &12:28:04.19 &12:28:04.20 &12:28:08.71 \\
$\delta(2000)$ &12:33:35.1 &12:33:06.5 &12:32:27.2 &12:34:25.7 \\
F606W (Vega System) &$22.86 \pm 0.01$ &$22.07 \pm 0.01$ &$24.59 \pm 0.02$
&$23.65 \pm 0.01$ \\
F814W (Vega System) &$21.92 \pm 0.01$ &$21.14 \pm 0.01$ &$23.79 \pm 0.02$
&$22.83 \pm 0.01$ \\
F606W$-$F814W &0.94 &0.93 &0.80 &0.82 \\
$V$ (transformed) &23.11 &22.32 &24.79 &23.86 \\
$V-I$ (transformed) &1.20 &1.19 &1.01 &1.04 \\
$V-I$ (dereddened) &1.16 &1.14 &0.97 &1.00 \\
Ellipticity ($b/a$)\tablenotemark{a} & 0.95$\pm$0.04 & 0.88$\pm$0.04 & 0.89$\pm$0.04 & 0.93$\pm$0.04\\
Estimated Luminosity ($10^5 L_{\odot}$) & 2.0 & 4.1 & 0.4 & 1.0 \\
Estimated Mass ($10^5 M_{\odot}$) & 3.1 & 6.5 & 0.7 & 1.6 \\
Half-light radius ($1\arcsec = 80$~pc) &$0 \farcs 025 \pm 0 \farcs 007$
&$0 \farcs 043 \pm 0 \farcs 005$ &  $0 \farcs 12$:
&$0 \farcs 10 \pm 0 \farcs 01$ \\
King Profile Core Radius &$0 \farcs 002 \pm 0 \farcs 001$
&$0 \farcs 005 \pm 0 \farcs 001$ &$0 \farcs 031 \pm 0 \farcs 012$
&$0 \farcs 037 \pm 0 \farcs 005$ \\
King Profile Tidal Radius &$1 \farcs 4 \pm 0 \farcs 2$
&$1 \farcs 5 \pm 0 \farcs 1$ & $1 \farcs 7$: &$1 \farcs 0 \pm 0 \farcs 2$ \\
King Profile $\chi^2/\nu$ & 25.6/17 &26.3/18 &0.96/10 &2.47/15 \\
$R^{1/4}$ Profile $\chi^2/\nu$ &26.8/18 &35.9/19 & 3.22/11 & 38.5/16 \\
Exp. Profile $\chi^2/\nu$ &226/18 &676/19 & 13.5/11 & 71.9/16 \\
\enddata
\tablenotetext{a}{Ellipticities were measured with the PyRAF task {\tt imexamine}.  Errors were calculated as $1-(b/a)$ for isolated stars in the image.}
\label{tab1}
\end{deluxetable}
\clearpage

\figcaption[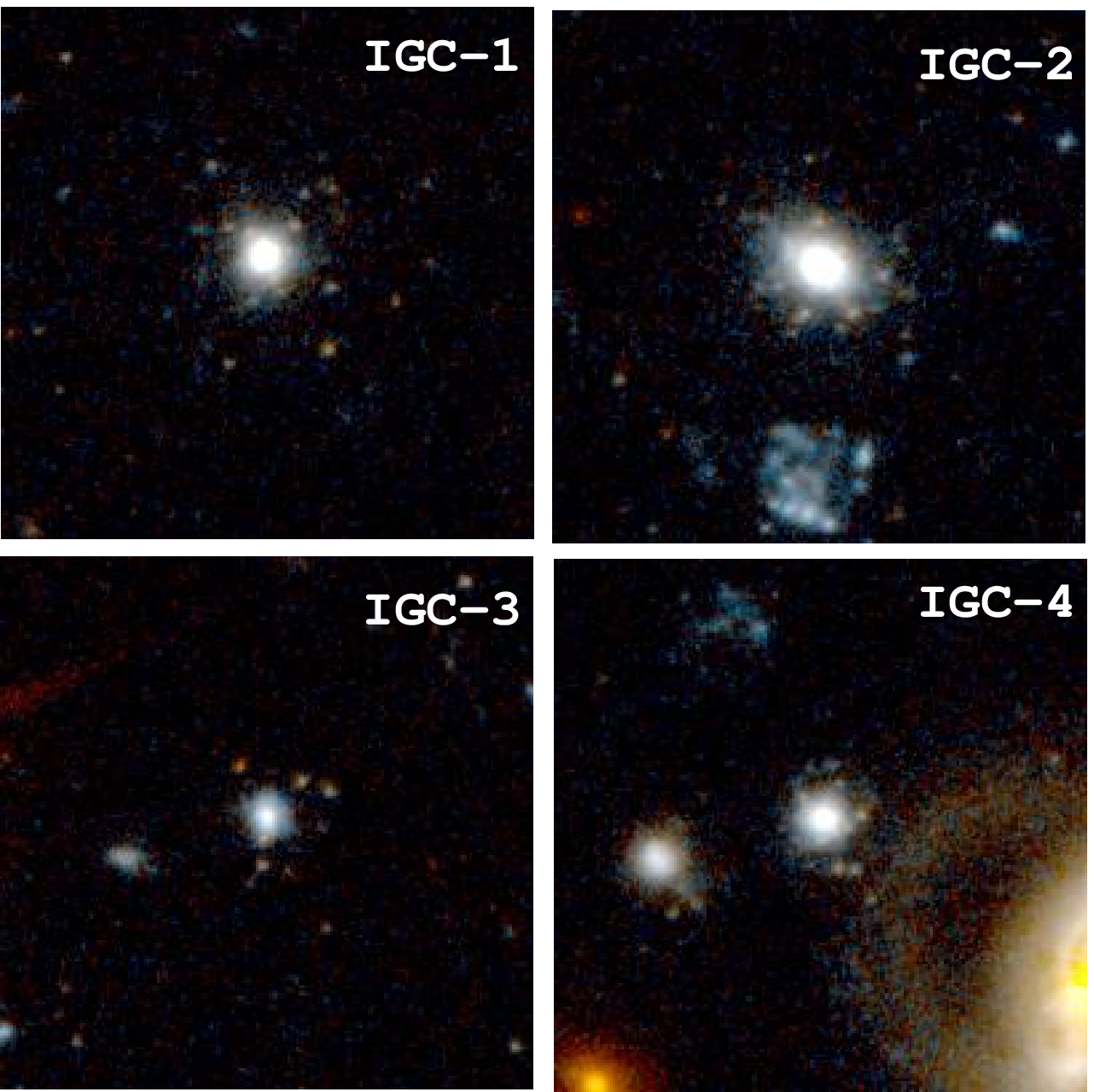]{Color images of our four IGC candidates produced
by combining our F606W and F814W exposures.  Each image is $6\arcsec
\times 6\arcsec$ on a side, with north up and east to the left.  Note
the number of point sources surrounding each candidate; these are
likely to be red giant stars in the outer regions of the clusters.
Also note the well-resolved background galaxies, for example, south of
IGC-2 and east of IGC-3 and IGC-4.  These objects are much more
extended than the stars or IGC candidates.
\label{pics}
}

\figcaption[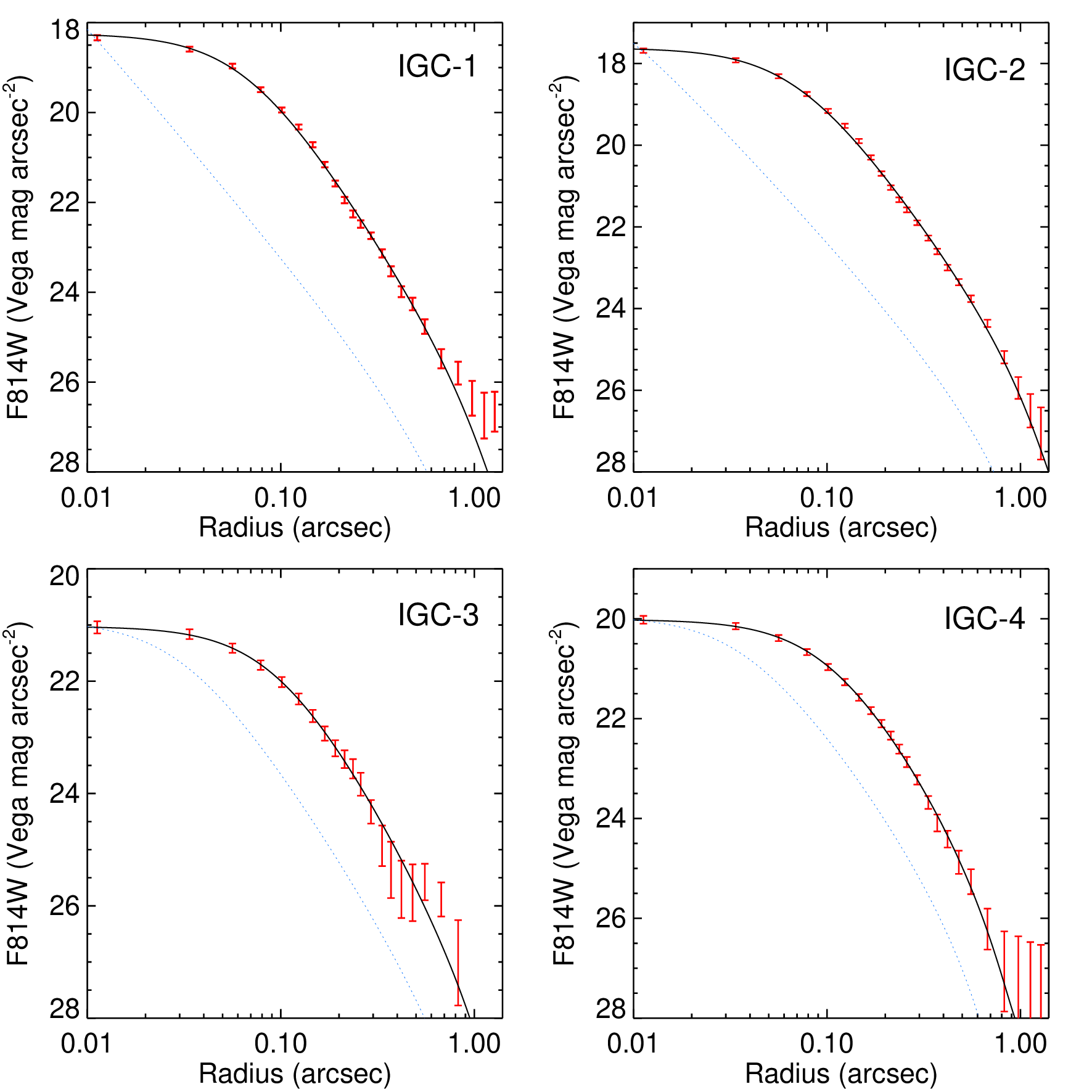]{The azimuthally averaged radial surface
brightness profiles of our 4 IGC candidates in the F814W filter.  The
dark solid curves show the best-fitting \citet{king62} profiles
convolved with a \citet{moffat} approximation of the {\sl ACS\/} point
spread function.  The unconvolved King profile is illustrated by the
light dotted line in each panel.  Fitting statistics and parameters
are given in Table~\ref{tab1}.  Note the excellent quality of the
fits; profiles based on \citet{deV} or exponential laws are not nearly
as good.
\label{king}
}

\figcaption[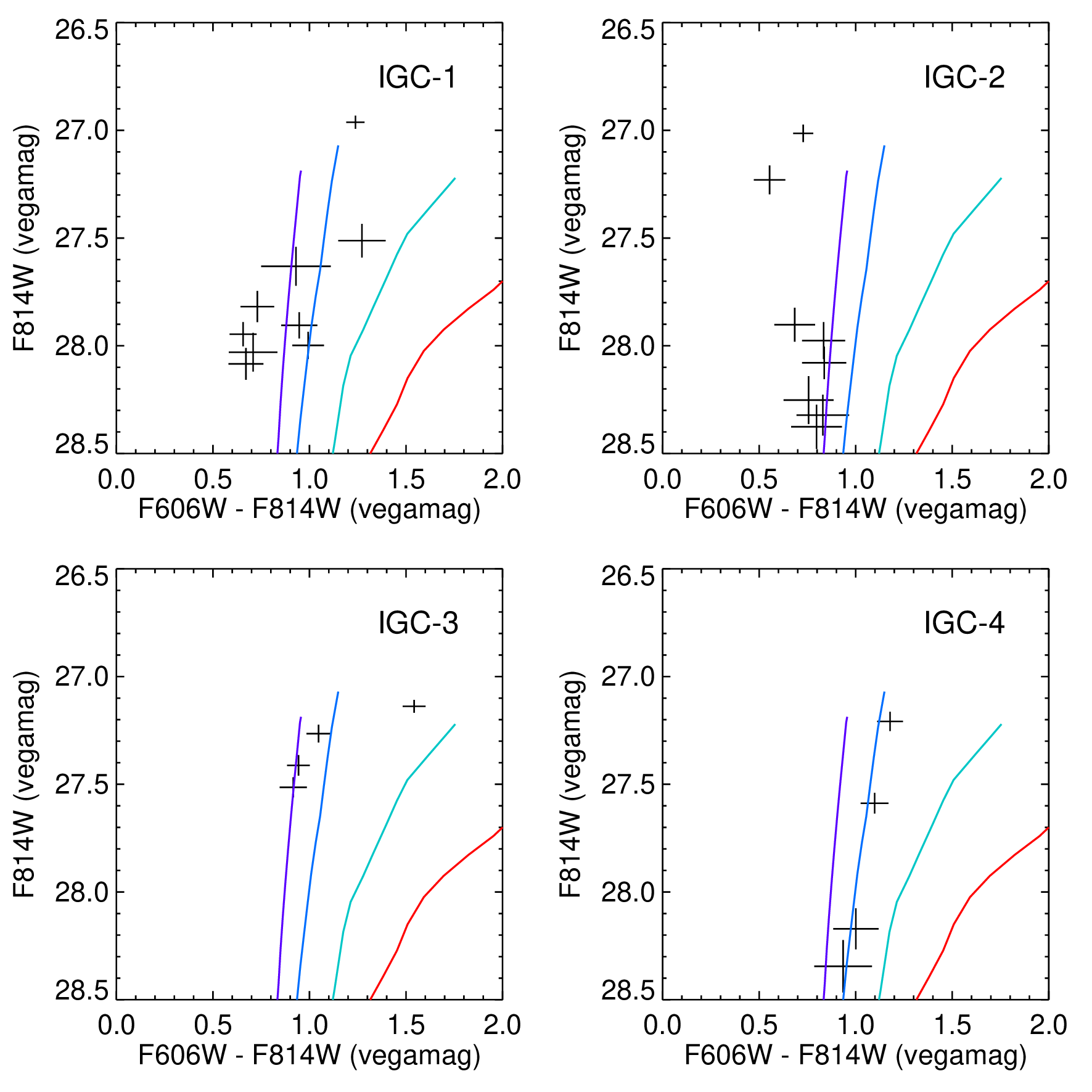]{The color-magnitude diagrams of the point sources
projected within $1\farcs 3$ of each IGC candidate.  Overplotted are
theoretical 12.5~Gyr isochrones for stellar populations with $Z =
0.008$ ([M/H]~$= -0.4$; red), $Z=0.004$ ([M/H]~$=-0.7$; cyan),
$Z=0.001$ ([M/H]~$=-1.3$; blue), and $Z=0.0001$ ([M/H]~$=-2.3$
(purple).  On average, only $\sim 0.67$ stars in each figure should be
a chance superposition.  Note that all four systems are metal-poor:
IGC-4, the most metal-rich of the group has [M/H] $\sim -1.3$.
\label{cmds}
}

\figcaption[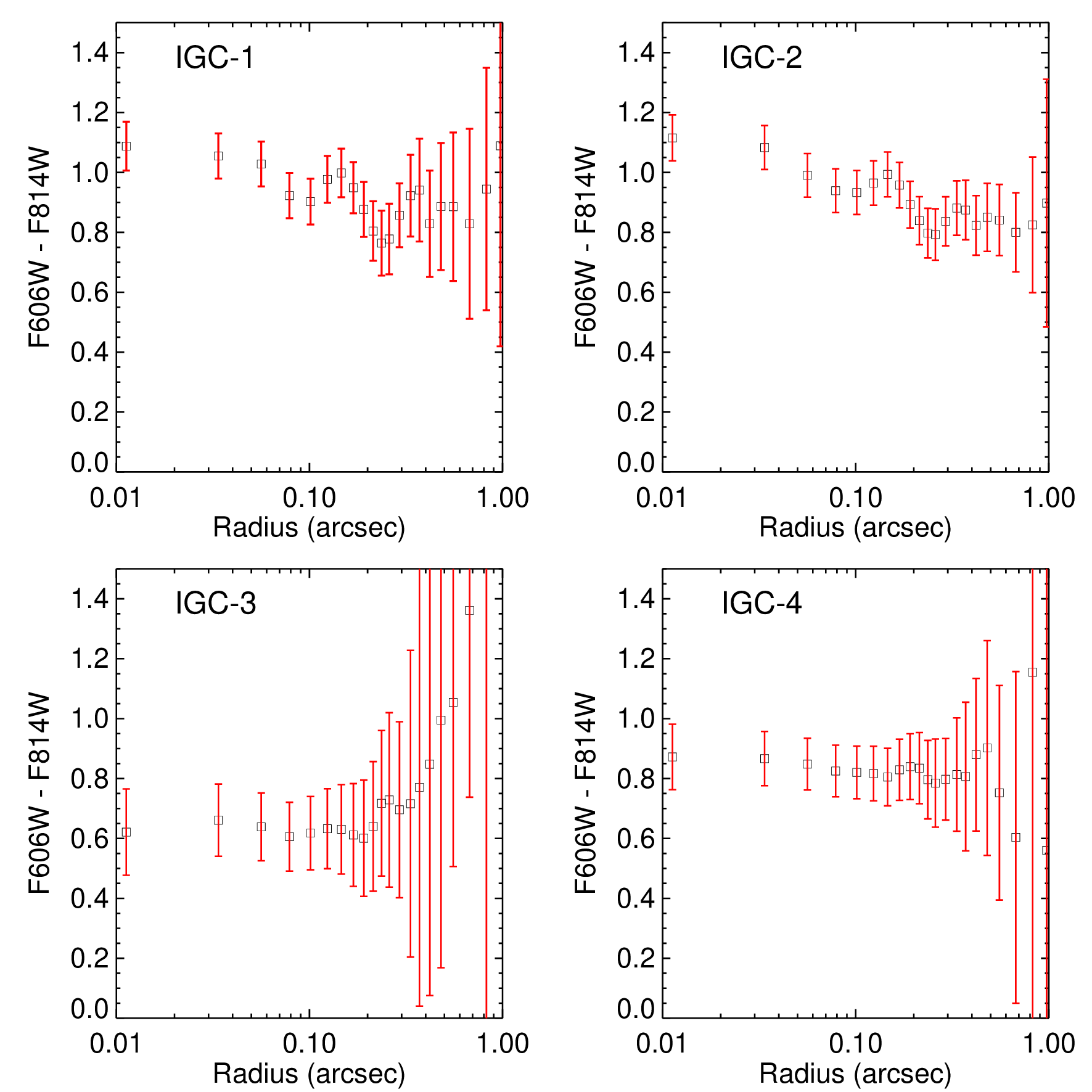]{The F606W$-$F814W colors of our globular cluster
candidates computed in a series of circular annuli.  Note the color
gradients associated with IGC-1 and IGC-2; roughly one-third of
Galactic clusters have such gradients.  This partially explains why
these metal-poor systems have integrated colors that are slightly redder
than the mean for M87 globulars.
\label{colorprof}
}

\figcaption[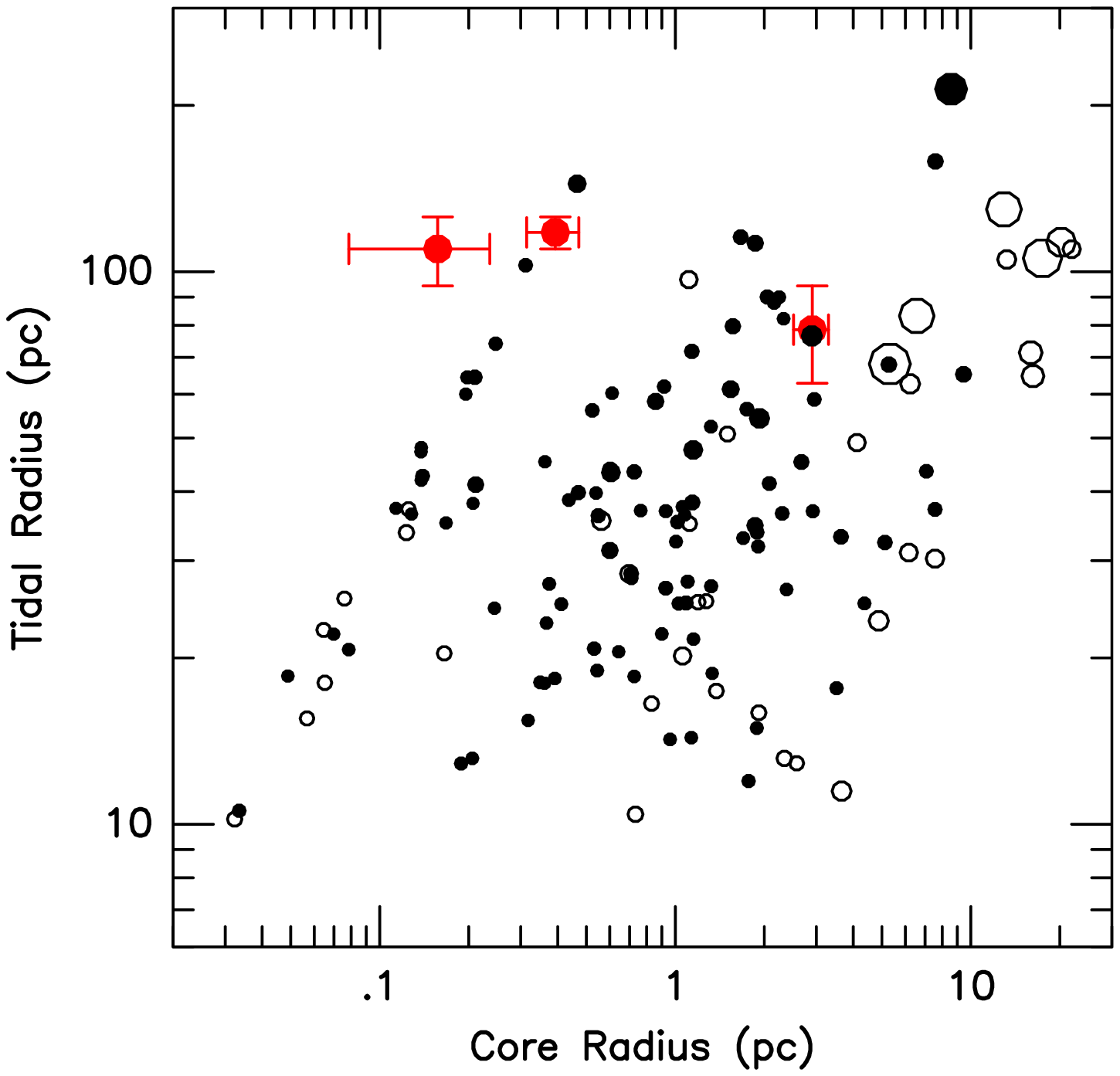]{The core radii and tidal radii of our IGC
candidates (plotted as large red points), excluding IGC-3 (whose tidal
radius is unconstrained).  For comparison, the parameters of Galactic
globular clusters are plotted in black points.  Solid black points are
clusters brighter than $M_V = -6.5$, which is the magnitude limit of
our observations.  The open circles represent fainter clusters.  The
sizes of the black points are proportional to their Galactocentric
distance (from 0.6 to 120 kpc).  Note that the intergalactic globulars
have systematically larger tidal radii than their Galactic
counterparts, and that many of the Galactic objects with large tidal
radii are at large Galactocentric distance.  This is consistent with
the idea that our IGC candidates have evolved in a largely stress-free
environment.
\label{rcrt}
}

\begin{figure}
\figurenum{1}
\plotone{f1.eps}
\end{figure}
\clearpage

\begin{figure}
\figurenum{2}
\plotone{f2.eps}
\end{figure}
\clearpage

\begin{figure}
\figurenum{3}
\plotone{f3.eps}
\end{figure}
\clearpage

\begin{figure}
\figurenum{4}
\plotone{f4.eps}
\end{figure}
\clearpage

\begin{figure}
\figurenum{5}
\plotone{f5.eps}
\end{figure}
\clearpage

\end{document}